\documentclass[range]{ar2e}
\usepackage{graphicx}
\usepackage{amsmath}
\usepackage{amssymb}

\begin{document}

\jname{Annu. Rev. Nucl. Part. Sci.}
\jyear{2010}
\jvol{60}
\ARinfo{}

\title{The Diffuse Supernova Neutrino Background}

\markboth{The Diffuse Supernova Neutrino Background}{John F. Beacom}

\author{John F. Beacom \\
\affiliation{
Department of Physics, \\
Department of Astronomy, \\
and Center for Cosmology and Astro-Particle Physics, \\
The Ohio State University, \\
191 West Woodruff Avenue, Columbus, Ohio 43210, USA \\
\medskip
beacom.7@osu.edu}
}


\begin{abstract}
The Diffuse Supernova Neutrino Background (DSNB) is the weak glow of MeV
neutrinos and antineutrinos from distant core-collapse supernovae. The DSNB
has not been detected yet, but the Super-Kamiokande (SK) 2003 upper limit
on the $\bar{\nu}_e$ flux is close to predictions, now quite precise, based on
astrophysical data. If SK is modified with dissolved gadolinium to reduce
detector backgrounds and increase the energy range for analysis, then it
should detect the DSNB at a rate of a few events per year, providing a new
probe of supernova neutrino emission and the cosmic core-collapse rate. 
If the DSNB is not detected, then new physics will be required. Neutrino
astronomy, while uniquely powerful, has proven extremely difficult --
only the Sun and the nearby Supernova 1987A have been detected to date
-- so the promise of detecting new sources soon is exciting indeed.
\newpage
\end{abstract}

\maketitle

\newpage


\section{Introduction}

The Diffuse Supernova Neutrino Background (DSNB) is the flux of
neutrinos and antineutrinos emitted by all core-collapse supernovae in
the causally-reachable universe; it will appear isotropic and
time-independent in feasible observations. (Hereafter, neutrinos means
neutrinos and antineutrinos, and supernovae means core-collapse
supernovae, unless specified.) It results from the quasi-thermal MeV
neutrino spectrum emitted by newly-formed neutron stars convolved with
the rapid redshift evolution of the supernova rate. The cosmic energy
density in neutrinos from core-collapse supernovae is comparable to that
in photons from stars, $\sim 0.01$ eV cm$^{-3}$, and is $\sim 10$
times less than that of the cosmic microwave background.  Some earlier
literature referred to ``supernova relic neutrinos," but ``relic" often
caused confusion with the $10^{-4}$ eV neutrinos from the Big Bang.

\medskip

Why is detection of the DSNB an important and relevant experimental goal?

\smallskip
$\bullet$
{\bf Understanding supernovae is crucial to astrophysics and physics.}
Supernovae are central to the cosmic history of stellar birth and death;
the production of chemical elements, neutron stars and black holes,
cosmic rays, and gravitational waves; and to exploiting the extreme
physical conditions of core collapse to probe the properties of
neutrinos and hypothetical particles.  New observations and syntheses
are needed.

\smallskip
$\bullet$
{\bf We cannot understand supernovae without detecting neutrinos.}
The optical supernova reveals little about the collapsing core itself,
which is embedded in the stellar envelope; the total energy in photons
is much less than that in neutrinos, leaves over a vastly longer time,
and depends on the properties of the progenitor star. Neutrinos are
emitted from the core, revealing its properties.

\smallskip
$\bullet$
{\bf Detecting bursts of neutrinos from nearby supernovae is difficult.}
Prodigious emission and large detectors partially compensate the tiny
detection cross section.  For Milky Way supernovae (distance $D \sim 10$
kpc; $1 {\rm\ pc} = 3.1 \times 10^{18} {\rm\ cm}$), a neutrino burst
would be detected easily, but the burst rate is only a few per century. 
For supernovae in nearby galaxies ($D \sim$ 1--10 Mpc), where the total
burst rate is above one per year, much larger detectors will be
required.

\smallskip
$\bullet$
{\bf The DSNB is a guaranteed steady source of supernova neutrinos.}
While the probability of detecting even a single neutrino from a distant
supernova is infinitesimally small, the number of supernovae per year is
astronomically large, yielding a nonzero rate. The total rate and
spectral shape of the DSNB are new probes of supernova neutrino emission
and the cosmic core-collapse rate.

\medskip

{\it The DSNB has not been detected yet, but the discovery prospects
are excellent.}
Super-Kamiokande (SK) is large enough to detect the DSNB at the level
of a few events per year.  These events are hidden by detector
backgrounds, which are already low and could be substantially reduced
with added gadolinium to detect neutrons, a proposed upgrade for which
there is active research and development.

Detection of the DSNB will yield science that is difficult to obtain
otherwise. It will measure the average supernova neutrino emission
spectrum, and comparison to burst detections will probe the star-to-star
variation.  It will also be sensitive to the optically-dark failed
supernovae expected from high-mass stars.

This review is written for experimentalists, observers, and theorists in
nuclear physics, particle physics, and astrophysics, and covers the many
aspects of the DSNB.  It is intended to be an accessible overview of the
main themes needed to predict, detect, and interpret the DSNB, with
details given in references.


\section{Impossible Dreams of Neutrino Astronomy}

Neutrino astronomy is a unique tool for answering some of the oldest and
most important questions in nuclear and particle
astrophysics~\cite{Freedman:2004rt}. What happens when a massive star
runs out of fuel?  Are observed high-energy gamma ray sources powered by
cosmic-ray protons or electrons?  What is the origin of the
highest-energy cosmic rays?  Are there astrophysical neutrino sources
unseen with electromagnetic observations?  Are there new properties of
neutrinos, or new particles, that can be revealed only with neutrino
detectors?

The immense discovery potential of neutrino astronomy arises from the
small interaction cross section of neutrinos with matter, which allows
neutrinos to escape from and thus to reveal the interiors of dense and
distant astrophysical objects, undeflected, undegraded, and unobscured. 
This powerful magic comes at a terrible price, as it is much less likely
that the neutrinos will interact with a detector than with an entire
star.  As far as we know, neutrinos interact only via the weak nuclear
force and gravity (but with only tiny masses).

In 1934, shortly after Pauli postulated the existence of the neutrino,
Bethe and Peierls wrote that ``If [there are no new forces] -- one can
conclude that there is no practically possible way of observing the
neutrino"~\cite{Bethe:1934qn}. It took until 1956 for neutrinos to be
detected, by Reines and Cowan, using a nuclear reactor
source~\cite{1956Sci...124..103C}, for which Reines shared in the 1995
Nobel Prize (Cowan died in 1974). This detection allowed a first
consideration of neutrino astronomy in principle.

Prescient remarks about this potential were made long ago, e.g.,
Refs.~\cite{1956Natur.178..446R, 1960Sci...131..299M,
1963SvPhU...6....1P, 1965Sci...147..115B, Ruderman:1965zz}. As Bahcall
wrote in 1964 about solar neutrinos, before their detection, ``Only
neutrinos, with their extremely small interaction cross sections, can
enable us to see into the interior of a star..."~\cite{Bahcall:1964gx}.
The detection of neutrinos from the Sun and the nearby Supernova 1987A
(SN 1987A) shows that the required sensitivity has begun to be reached.
Davis and Koshiba shared in the 2002 Nobel Prize for these first
detections in neutrino astronomy, which led a precise understanding of
the solar fusion reactions, the discovery of neutrino mixing, and
confirmation that Type II supernovae result from the deaths of massive
stars.

No other astrophysical neutrinos have been detected, despite decades of
effort.  In Bahcall's 1989 book, {\it Neutrino Astrophysics}, which was
almost completely about solar neutrinos, he wrote that ``The title is
more of an expression of hope than a description of the book's
contents....the observational horizon of neutrino astrophysics may grow
... perhaps in a time as short as one or two
decades"~\cite{1989neas.book.....B}. Indeed, the first detections of
distant neutrino sources now seem imminent, due to more certain
predictions, based on new astronomical and laboratory data, and due to
astounding improvements in detector sensitivity.  Any detections will be
of enormous scientific importance, certain to provide new information.

This review is about the exciting prospects for new results in
extrasolar MeV neutrino astronomy; as described below, these build on a
long history of work in experiment, observation, and theory. There are
similarly encouraging prospects at the TeV ($10^6$ MeV)
scale~\cite{Learned:2000sw, Anchordoqui:2009nf} and the EeV ($10^{12}$
MeV) scale~\cite{Beatty:2009zz, Anchordoqui:2009nf}.


\section{Framework: DSNB Detection Spectrum at Earth}

Here the framework to calculate the DSNB detection spectrum is given,
along with a summary of its three ingredients: the supernova neutrino
emission (Section 4), the cosmic supernova rate (Section 5), and the
detector capabilities (Section 6), followed by a return to the framework
(Section 7), then SK observations (Section 8), background reduction
(Section 9), and conclusions (Section 10).


\subsection{Simple Estimate of the Detection Rate}

SN 1987A was discovered optically in the Large Magellanic Cloud, a
nearby satellite of the Milky Way, at a distance of about 50
kpc~\cite{1989ARA&A..27..629A}. Since it was so close, it was likely
that its neutrinos had interacted in the Kamiokande-II (Kam-II) and
Irvine-Michigan-Brookhaven (IMB) water-\v{C}erenkov detectors,
originally intended for proton decay searches. Kam-II reported 12 events
and IMB reported 8 events, with energies of several tens of MeV, in
contemporaneous bursts of about 10 seconds, occurring a few hours before
the optical brightening~\cite{Hirata:1987hu, Hirata:1988ad,
Bionta:1987qt, Bratton:1988ww}. The events were largely consistent with
being due to inverse beta decay, $\bar{\nu}_e + p \rightarrow e^+ + n$,
where the positron is detected, and where its energy is close to that of
the neutrino.

Deferring many details, we can simply estimate the (steady) detection
rate of DSNB neutrinos from the (burst) detection rate of SN 1987A
neutrinos, as
\begin{equation}
\left[\frac{dN_\nu}{dt} \right]_{DSNB}
\sim \,
\left[\frac{dN_\nu}{dt} \right]_{87A}
\left[\frac{N_{SN} \, M_{det}}{4\pi D^2}\right]_{87A}^{-1} \;
\left[\frac{N_{SN} \, M_{det}}{4\pi D^2}\right]_{DSNB} \,.
\label{eq:conversion}
\end{equation}
The detection rate in Kam-II during the burst was $\sim 1$ s$^{-1}$. 
How are the inputs for the dimensionless product of conversion factors
related?  For SN 1987A, $N_{SN} = 1$, while there are $N_{SN} \sim 100$
neutrino bursts in the universe starting within any 10-second interval;
this accounts for the supernova rate of the Milky Way, $\sim 10^{-2}$
year$^{-1}$, the space density of comparable galaxies, $\sim 10^{-2}$
Mpc$^{-3}$, and the fact that the supernova rate per galaxy was $\sim
10$ times higher in the past.  The detector mass $M_{det}$ of Kam-II was
$\sim 10$ times smaller than that of SK, which is 22.5 kton ($22.5
\times 10^9$ g). The distance $D$ of SN 1987A was 0.050 Mpc, whereas for
a typical supernova contributing the DSNB, at $z \sim 1$, it is $c / H_0
\sim 4000$ Mpc, $\sim 10^5$ times farther.  Then the time-averaged DSNB
detection rate in SK, noting changes in $N_{SN}$, $M_{det}$, and
$1/D^2$, respectively, is
\begin{equation}
\left[\frac{dN_\nu}{dt} \right]_{DSNB}
\sim \, 
(1 {\rm\ s}^{-1}) \times 100 \times 10 \times 10^{-10}
\, \sim \, 10^{-7} {\rm\ s}^{-1}
\, \sim \, 3 {\rm\ year}^{-1} \,,
\label{eq:estimate}
\end{equation}
which is about the right number.

The Poisson expectation for the number of neutrinos detected per
supernova is $\ll 1$, so nearly all supernovae will yield 0, and a rare
few unidentified supernovae will yield just 1. While the long-term
average of the neutrino detection rate from Milky Way supernovae is
$\sim 100$ times larger than for the DSNB, it requires waiting decades
for large bursts from single identified supernovae. Future very large
detectors may detect more frequent, but much smaller, bursts from
identified supernovae in the nearest galaxies. Burst detection will be
required to study the time profile and flavor dependence of supernova
neutrino emission.


\subsection{Line of Sight Integral for the DSNB Flux}

The DSNB flux spectrum at Earth is calculated from the neutrino emission
per supernova $\varphi(E_\nu)$ and the evolving core-collapse
supernova rate $R_{SN}(z)$ as
\begin{equation}
\frac{d\phi}{dE_\nu}(E_\nu) \, = \,
\int_0^\infty \left[(1+z) \phantom{\frac{a}{b}} \!\!\! \varphi[E_\nu (1 + z)] \right]
\left[R_{SN}(z) \phantom{\frac{a}{b}} \!\!\! \right]
\left[\left| \frac{c \, dt}{dz} \right| dz\right] \,,
\label{eq:fluxspectrum}
\end{equation}
where this form arises as a standard line-of-sight integral for the
radiation intensity (flux per solid angle) from a distribution of
sources. Neglecting cosmological factors, the differential injection
rate of neutrinos depends on the rate density (the number of neutrinos
emitted per volume and time, i.e., the supernova neutrino emission times
the supernova rate density) times the volume element $d \Omega \, r^2
dr$, while the fluxes fall off as $1/(4 \pi r^2)$, canceling the
$r^2$. The received intensity is then just a line integral of the rate
density.  Since the sources are isotropic on the sky, the detection
reaction has little directionality, and Earth is transparent to
neutrinos, the intensity has been integrated over $4 \pi$ to obtain the
flux.

This defines the neutrino number flux spectrum, in units cm$^{-2}$
s$^{-1}$ MeV$^{-1}$.  To obtain the energy flux spectrum, more familiar
to photon astronomers, multiply through by $E_\nu$. The first term in
large brackets is the emission spectrum $\varphi(E_\nu)$, where a
neutrino received at energy $E_\nu$ was emitted at a higher energy
$E_\nu (1 + z)$; the prefactor of $(1 + z)$ on the spectrum accounts for
the compression of the energy scale.  The second term is the supernova
rate density $R_{SN}(z)$. The third term is the differential distance,
where $| \mathrm{d}t / \mathrm{d}z |^{-1} = H_0 (1+z) [\Omega_\Lambda +
\Omega_m(1+z)^3]^{1/2}$, the cosmological parameters are $H_0 = 70$ km
s$^{-1}$ Mpc$^{-1}$, $\Omega_m = 0.3$, $\Omega_\Lambda = 0.7$, and their
uncertainties can be neglected.

Above, it is assumed that there is no variation in the supernova
neutrino emission among different progenitor stars, which follows from
the core-collapse criterion for massive stars and how supernova neutrino
mixing effects work, both discussed in Section 4.2. It is also assumed
that nothing except redshifting happens to the neutrinos between the
supernovae and Earth, as obscuration is always negligible, the neutrino
mass eigenstates that propagate do not oscillate, and no exotic effects
are expected (but such can be tested~\cite{Ando:2003ie, Fogli:2004gy,
Goldberg:2005yw, Baker:2006gm}). These are realistic assumptions at the
precision of the expected counts and energy resolution, and can be
generalized with appropriate averages.


\begin{figure}[t]
\centerline{\includegraphics[width=4in]{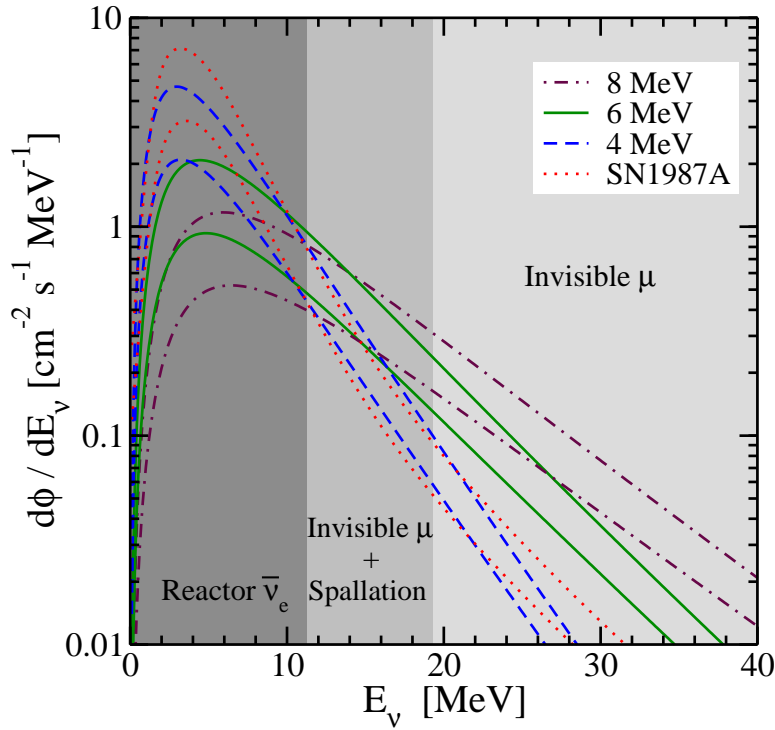}}
\caption{\label{fig:fluxspectrum}
{\bf Predicted DSNB $\bar{\nu}_e$ flux spectrum.}
The spread between different labeled temperature bands indicates the
uncertainty in the supernova neutrino emission; the widths of the bands
indicates the uncertainty in the cosmic supernova rate.  Grey shaded
regions indicate energy ranges where noted detector backgrounds are
important. Figure as shown in Ref.~\cite{Horiuchi:2008jz}.
}
\end{figure}

\subsection{Contemporary Inputs, Uncertainties, and Detectors}

Throughout, we focus on the effective $\bar{\nu}_e$ spectrum at the
surface of the supernovae, after any neutrino mixing effects in the
star, as this is the same observable for both burst and DSNB detection.
As described in Section 4, the neutrino emission in one flavor is
characterized by its total energy (time-integrated luminosity) and
average energy ($3.15 \, T$ for a Fermi-Dirac spectrum, where $T$ is the
temperature in units of MeV). These are the parameters to be determined
by a DSNB measurement, as the supernova neutrino emission is now the
most uncertain input, and it cannot be measured by astronomical
techniques.  Nominal values are $E_{\bar{\nu}_e, tot} \sim 5 \times
10^{52}$ erg ($3 \times 10^{58}$ MeV), and $E_{\bar{\nu}_e, avg} \sim
15$ MeV. If a black hole forms promptly, the luminosity from the
collapsing core prior to being abruptly terminated can be higher than
for neutron star formation, yielding a larger time-integrated
emission~\cite{2008PhRvD..78h3014N}.

The cosmic supernova rate is now well measured, as shown in Section 5.
For the supernova rate density, the local value is $\simeq (1.25 \pm
0.25) \times 10^{-4}$ Mpc$^{-3}$ year$^{-1}$, and the rate grows by a
factor $\simeq 10$ by $z = 1$. DSNB measurements are also sensitive to
core collapses that are optically dark but neutrino bright.
 
{\bf Figure~\ref{fig:fluxspectrum}} shows the $\bar{\nu}_e$ flux
spectrum; the SK 2003 limit is $\lesssim 1.2$ cm$^{-2}$ s$^{-1}$ for
energies integrated above 19.3 MeV, ruling out spectra above about
the middle of the $T = 8$ MeV band. Since the predictions are for
generic thermal spectra, they could be used for other flavors as well.
Equation~(\ref{eq:fluxspectrum}) defines the {\it flux spectrum} of
neutrinos. In Section 6, the {\it event rate spectrum} of positrons from
the reaction $\bar{\nu}_e + p \rightarrow e^+ + n$ is introduced to
allow comparison with the SK data.


\subsection{Theoretical Predictions and Experimental Revolutions}

Theoretical predictions of the DSNB flux go back more than forty years,
e.g., to brief mentions by Zel'dovich and
Guseinov~\cite{1965SPhD...10..524Z} and Ruderman~\cite{Ruderman:1965zz},
and a calculation by Guseinov~\cite{1967SvA....10..613G}.  First examples
of detailed calculations, from the early 1980s, are
Refs.~\cite{1982SvA....26..132B, 1984SvA....28...30D,
1984Natur.310..191K, 1984NYASA.422..319B,
1985Natur.316..420L, Dar:1984aj, 1986ApJ...302...19W,
1986NCimC...9..443S}; unfortunately, the true flux is not as large as in
these predictions (but see the pessimistic lower limit noted in
Ref.~\cite{1986ApJ...302...19W}). The variation between predictions,
formerly orders of magnitude, became much less beginning in the 
mid-1990s, in e.g., Refs.~\cite{Totani:1995dw, Malaney:1996ar, 
Hartmann:1997qe, Kaplinghat:1999xi, Ando:2002ky, Fukugita:2002qw, 
Strigari:2003ig, Iocco:2004wd, Strigari:2005hu, Daigne:2005xi, 
Lunardini:2005jf}.  At roughly the end of that era, the DSNB was reviewed
in 2004 by Ando and Sato~\cite{Ando:2004hc}.  The SK 2003 upper limit on
the DSNB was a key development, as it directly rules out models with fluxes
more than a few times larger than present predictions.

It would be unfair to judge earlier predictions without adjusting their
inputs to contemporary values. This seems difficult, as a wide variety
of different formalisms, inputs, standards of precision, and methods of
quoting results were used. However, this is easy in the present
perspective, first noted in Ref.~\cite{Yuksel:2005ae}, where the
principal remaining unknown is the supernova neutrino emission, and the
assumed values could be easily extracted from earlier papers. In fact,
most DSNB predictions did use reasonable supernova neutrino emission
parameters (for the earliest papers, the existence of $\nu_\tau$ and
$\bar{\nu}_\tau$ was unknown), even before the detection of SN
1987A. There were huge differences in the supernova rate density
assumed, due to uncertainties that have been dramatically reduced by
comprehensive star formation and supernova rate measurements over the
past decade. The results shown in this review cover the range of
reasonable allowed models for both the supernova neutrino emission and
the supernova rate density.


\section{First Ingredient: Supernova Neutrino Emission}

The detection of neutrinos from SN 1987A established that {\it some}
supernovae produce {\it some} neutrinos. Of course, our knowledge is
much greater than that -- there is strong evidence that core-collapse
supernovae can be identified optically, and that a powerful neutrino
burst must accompany core collapse.

To understand the properties of massive stars and how supernovae
explode, it is necessary to confirm the ubiquity of neutrino emission
from all types of supposed core-collapse supernovae and to refine our
knowledge of this emission.


\subsection{The Fates of Massive Stars}

Massive stars race towards their doom. The lifetime of a star, mostly
the time spent burning hydrogen in the core, is $\sim 30 \, (8 M_\odot
/ M)^{2.5} \times 10^6$ year, where $M_\odot = 2.0 \times 10^{33}$ g is
the solar mass~\cite{1994sipp.book.....H}. The fusion reactions proceed
in stages that depend on temperature, overcoming the Coulomb barriers
for successively heavier elements only once the preceding elements are
exhausted and the core contracts.  Only stars with mass greater than about
$8 M_\odot$ will become hot and dense enough in the core to burn past
carbon and oxygen to iron~\cite{1984ApJ...277..361K, Heger:2002by}. A
Chandrasekhar instability sets in once about $1.4 M_\odot$ of iron has
been created, since no further nuclear reactions can generate energy and
electron degeneracy pressure provides insufficient support. 
Photo-nuclear reactions then endothermically destroy iron, leading to
further collapse, which continues until the core reaches nuclear
densities and cannot be compressed further.  The subsequent bounce leads
to the formation of an outgoing shock, which removes the envelope of the
star and causes the optical supernova, which is bright for several
months.  What will be left behind is a neutron star or black hole from
the core and a shell remnant from the envelope.

Astronomical characterizations of supernovae are based on their optical
properties~\cite{Filippenko:1997ub}. Supernovae must be distinguished by
their spectra, as their light curves alone are not usually enough. Many
lines of evidence indicate that Types II, Ib, and Ic supernovae arise
from the core collapse of massive stars, differing in the amount of
progenitor envelope mass loss before explosion. Type II supernovae are
several times more common than Types Ib and Ic, and the latter two seem
to arise from ranges of successively larger progenitor mass.  All should
have comparable emission of all flavors of neutrinos and antineutrinos,
due to the mechanism of core collapse.  Type Ia supernovae, which are
also several times less common than Type II supernovae, seem to be
powered by the runaway thermonuclear burning of a carbon and oxygen
white dwarf in a binary system into iron; only relatively modest emission
of low-energy electron neutrinos is expected.  Detecting neutrinos will test
the mapping between optical supernova types and underlying physical
mechanisms~\cite{Kistler:2008us}.


\subsection{Core-Collapse Supernova Neutrino Emission}

The basic picture of core collapse is well
known~\cite{1996slfp.book.....R}. When the core collapses, there is a
huge change in its gravitational potential energy.  For a constant
density profile, the change in this self-energy is
\begin{equation}
\Delta(P.E.) \, \simeq \, 
\left(\frac{3 G_N M^2}{5 R}\right)_{NS} -
\left(\frac{3 G_N M^2}{5 R}\right)_{core} 
\simeq 
3 \times 10^{53} {\rm\ erg} \,,
\label{eq:totalenergy}
\end{equation}
where the number follows from the mass and radius of a typical neutron
star, $1.4 \, M_\odot$ and 10 km (the subtracted term is negligible
since the initial core radius is thousands of kilometers), and this is
an incredible $\sim 10\%$ of $M c^2$ for the
core~\cite{2001PhRvL..86.5647H, Page:2006ud, Lattimer:2006xb}.  The
mass of the collapsed core varies little with the mass of the progenitor
star~\cite{2008ApJ...679..639Z, 2009ApJ...699..409F}. Since the
proto-neutron star is near nuclear density, no particles except
neutrinos can escape, and core collapse is inevitably accompanied by a
huge burst of neutrinos, carrying away nearly the whole change in
gravitational potential energy. The neutrinos, which begin leaving the
core at the instant of collapse, arrive at Earth before the light, which
is delayed by at least the time it takes the shock to propagate through
the envelope, typically hours.  This was observed for SN 1987A,
and is a crucial confirmation of the basic picture.

It is thought that each of the six flavors of neutrinos and
antineutrinos carries away a comparable fraction of the total energy. 
In numerical simulations of supernovae, this equipartition is only
approximate, and the sense in which it might be violated in nature is
not known~\cite{Mezzacappa:2005ju, Ott:2008jb, Janka:2006fh,
Huedepohl:2009wh}. Most of the neutrino emission is in pairs,
$\bar{\nu}_e + \nu_e$, $\bar{\nu}_\mu + \nu_\mu$, and $\bar{\nu}_\tau +
\nu_\tau$, that are produced by weak interactions in the hot and dense
matter in the core. The reaction $e^- + p \rightarrow \nu_e + n$
converts the nearly equal numbers of neutrons and protons from nuclei
into the neutron-rich matter of the neutron star, producing an excess
$\nu_e$ flux, but this is a $\sim 10\%$ effect.

The proto-neutron star density is so high that even neutrinos do not
readily escape, and instead diffuse out over several seconds, and are
thus emitted with a quasi-thermal spectrum characteristic of the surface
of last scattering. If the proto-neutron star were semi-transparent, the
neutrinos would leave on a timescale of $\sim 1$ millisecond, with
energies of $\sim 100$ MeV.  Taking opacity into account, and assuming
blackbody emission, $L \sim E_{tot}/\tau \sim R^2 T^4$, the increase in
the emission time by $\sim 10^4$ corresponds to the decrease in neutrino
average energy by $\sim 10$. A long emission timescale and moderate
temperature were observed for SN 1987A, confirming the high density of
the proto-neutron star.

\begin{figure}[t]
\centerline{\includegraphics[width=4in]{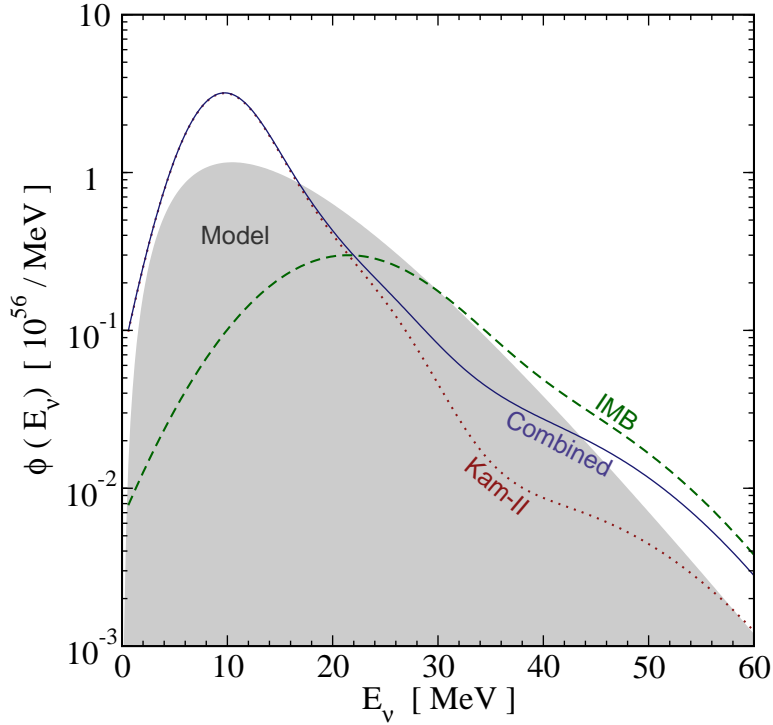}}
\caption{\label{fig:87Aspectrum}
{\bf Time-integrated effective $\bar{\nu}_e$ spectrum from SN 1987A.}
The curves indicate results reconstructed using only the Kam-II data,
the IMB data, or their combination.  The grey shaded region is a thermal
spectrum with parameters $E_{\bar{\nu}_e, tot} \sim 5 \times 10^{52}$
erg and $E_{\bar{\nu}_e, avg} \sim 15$ MeV.  Figure as shown in
Ref.~\cite{Yuksel:2007mn}.
}
\end{figure}

The opacities of neutrino flavors with matter vary, leading to different
neutrino temperatures, since the matter temperature of the proto-neutron
star falls with radius.  The interactions of $\nu_\mu$, $\nu_\tau$,
$\bar{\nu}_\mu$, $\bar{\nu}_\tau$ are only through the weak neutral
current, since their energies are too low to produce the corresponding
charged leptons.  For $\nu_e$ and $\bar{\nu}_e$, there are also
interactions through the weak charged current, so the cross sections are
larger; for $\nu_e$, the coupling constants are larger and there are
more neutron than proton targets.  Oft-quoted estimates for the
temperatures are $T = 8$ MeV for $\nu_\mu$, $\nu_\tau$, $\bar{\nu}_\mu$,
$\bar{\nu}_\tau$, $T = 5$ MeV for $\bar{\nu}_e$, and $T = 4$ MeV for
$\nu_e$.  More recent theoretical work indicates that the temperatures,
and the splittings between them, are less~\cite{Keil:2002in}.

Neutrino mixing effects relate the initial neutrino spectra at the
opaque emitting surface of the proto-neutron star to those outside the
supernova.  While the opacity is low outside the neutron star, the high
density of matter and even other neutrinos significantly affects
neutrino mixing~\cite{Dighe:1999bi, Fetter:2002xx, Duan:2006jv,
Dasgupta:2009mg}. The passage through high densities leaves the
neutrinos in incoherent mass eigenstates that do not oscillate between
the supernova and Earth. Mixing effects in Earth are suppressed by the
isotropy of the sources and the detection cross
section~\cite{Ando:2002zj}.

Neither the expected initial spectra nor the neutrino mixing effects are
understood well enough, so we use the observable effective spectra. For
$\bar{\nu}_e$, the effective time-integrated spectrum after mixing,
assumed to approximately be of the Fermi-Dirac form, in units
MeV$^{-1}$, is
\begin{equation}
\varphi(E_\nu) \, = \,
E_{\bar{\nu}_e, tot} \, \frac{120}{7 \pi^4} \, \frac{E_\nu^2}{T^4} \,
\frac{1}{(e^{E_\nu/T} + 1)} \,,
\label{eq:FDspectrum}
\end{equation}
where the total energy is $E_{\bar{\nu}_e, tot}$ and the average energy
is $E_{\bar{\nu}_e, avg} = 3.15 \, T$.  These parameters are not fixed
from theory, but will be determined from experiment, and will test
supernova simulations and neutrino mixing scenarios.


\subsection{Comparison to Supernova 1987A}

While the SN 1987A data were sparse, and the detectors were not
optimized for that purpose, these data are the most direct evidence we
have about supernova neutrino emission. A reconstruction of the
effective $\bar{\nu}_e$ spectrum from the Kam-II and IMB data is shown
in {\bf Figure~\ref{fig:87Aspectrum}}; it is in reasonable agreement
with expectations. Higher temperatures are favored by theoretical models
including neutrino mixing, as well as by observations of nucleosynthesis
yields that depend on high-energy neutrinos~\cite{Woosley:1988ip,
Heger:2003mm, Yoshida:2005uy}. It remains to be seen if SN 1987A was
different from an average supernova, and more data are needed,
especially at high energies.

The SN 1987A data are an important input for estimating the DSNB
spectrum~\cite{Fukugita:2002qw, Lunardini:2005jf}, though care must be
taken to note that the high-energy SN 1987A neutrinos are the most
relevant for predicting the DSNB~\cite{Yuksel:2007mn}. The SK 2003
search used a positron energy threshold of 18 MeV, and future searches
may reach 10 MeV; redshift effects enhance the importance of high-energy
emission.


\section{Second Ingredient: Cosmic Supernova Rate}

Supernovae are infrequent in the Milky Way, but not in the universe.
Since massive stars have lifetimes that are very short on cosmological
timescales, their cosmic birth and death rates are exactly equal.  The
cosmic star formation rate has been measured precisely, using a great
variety of techniques, principally based on the emission of massive
stars. This is enough to accurately give the rate of core collapses, and
it is supplemented by direct measurements of the optical supernova rate,
which have lower precision, but are in good agreement.


\subsection{Measured Star Formation Rate}

Star formation rate measurements do not literally probe star {\it
formation}, but begin with the total luminosity of the stars in a
galaxy. If these stars were all of the same known mass, so that their
luminosity and lifetime were given by stellar evolution theory, then the
star formation rate would be determined by the total mass of these stars
divided by their lifetime; this is the birth (and death) rate needed to
keep the number of stars in equilibrium. The main star-formation rate
indicators measure the emission from massive stars that are short-lived
on galactic timescales, so the equilibrium assumption should be good on
average.

A wide range of stellar masses are always present, complicating this
procedure, but the degeneracies can be broken, since stars of increasing
mass have higher temperatures and much higher luminosities. Massive
stars can be isolated by measurements of their continuum flux or of
nebular emission lines from atomic recombinations in the surrounding gas
they ionize. The distribution of stellar masses formed when star
formation occurs is assumed to follow a universal initial mass function
(IMF), e.g., the conventional Salpeter IMF scales as $\psi(M) = dn/dM
\propto M^{-2.35}$ for stellar masses $M$ between $0.1 M_\odot$ and $100
M_\odot$~\cite{1955ApJ...121..161S, 2003ApJ...593..258B}; more realistic
IMFs modify the low-mass end. With an IMF, calculated stellar and
nebular emission spectra, and careful consideration of the effects of
obscuration and re-radiation, the star formation rate can be determined
from a galaxy luminosity spectrum~\cite{Kennicutt:1998zb}. The high-mass
star formation rate is {\it measured}, but the formation rate of low-mass
stars has to be {\it deduced} assuming an IMF.  This introduces a factor-two
range in the normalization of the total star formation rate, due to the
low-mass IMF uncertainties.

The comoving star formation rate density, in units $M_\odot$ Mpc$^{-3}$
year$^{-1}$, is about ten times larger at $z = 1$ than it is today, at
$z = 0$. Comoving coordinates are those for which the expansion of the
universe has been removed; crudely, this defines the star formation rate
per average galaxy. Aside from the normalization uncertainty due to
choice of IMF, the precision of the combined data is at the
tens-of-percent level for $z \lesssim 1$~\cite{Hopkins:2006bw,
Horiuchi:2008jz}.  From $z = 1$ to at least $z \sim 4-5$, the star
formation rate is nearly flat, with a precision becoming as poor as a
factor of two, until a more uncertain decline at higher redshifts, e.g.,
Refs.~\cite{Yuksel:2008cu, Kistler:2009mv, Butler:2009nx, Oesch:2009bf,
Bouwens:2009at, Yan:2009qa}.

These results, especially those in the $z \lesssim$ 1--2 range relevant
for the DSNB, are quite robust, since they have been measured by many
different groups using different instruments and techniques, and
extensive efforts have been made to calibrate the observables and their
corrections for dust obscuration and incompleteness. The star formation
rate data are also in good agreement with measurements of the
extragalactic background light, which, analogous to the DSNB, is an
integral measure of stellar emission over redshifts~\cite{Horiuchi:2008jz}.
These and other data favor an IMF of somewhat shallower slope than the
Salpeter IMF.


\subsection{Predicted Supernova Rate}

To predict the core-collapse supernova rate density, $R_{SN}(z)$, in
units Mpc$^{-3}$ year$^{-1}$, from the star formation rate density,
$R_{SF}(z)$, we use
\begin{equation}
R_{SN}(z) \, = \, R_{SF}(z) \; 
\frac{\int_{8}^{50}\psi(M)dM}{\int_{0.1}^{100} M \psi(M)dM}
\, \simeq \,
\frac{R_{SF}(z)}{143 M_\odot} \,,
\label{eq:SNrate}
\end{equation}
where the upper integral gives the number of stars that lead to core
collapse and the lower integral gives the total mass in stars, and a
Salpeter IMF has been used.  Importantly, for $R_{SN}(z)$ the
above-mentioned normalization uncertainty due to the choice of IMF goes
away, because nearly the same massive stars source the star formation
rate measurements and lead to core-collapse supernovae.  For an IMF with
a shallower slope, $R_{SF}(z)$ is smaller, but the ratio of integrals is
larger, canceling in $R_{SN}(z)$ to the few-percent
level~\cite{Horiuchi:2008jz}.

In the upper integral, the choice of upper limit of integration is
unimportant, as long as it is large; then the integral scales as $(8
M_\odot / M_{lower})^{1.35}$ for a Salpeter IMF. Theoretical and
indirect empirical work indicates that the minimum mass star that leads
to core collapse is about $8 M_\odot$.  This is now supported by direct
observations. Using archival imaging from the Hubble Space Telescope and
other sources, it has been possible to identify the stellar progenitors
of some supernovae in nearby galaxies, e.g., Refs.~\cite{GalYam:2006iy,
Li:2007jj, Prieto:2008bw, Smartt:2009zr}. The luminosity of the star
determines its mass, and a fit to progenitor data gives a minimum mass
to lead to a core-collapse supernova of $(8.5 \pm 1.5) \,
M_\odot$~\cite{Smartt:2008zd}. In addition, the largest estimate so far
of the maximum mass star that ends instead as a white dwarf is $\sim 7
M_\odot$~\cite{2008ApJ...676..594K, 2009ApJ...693..355W}.

There is therefore a precise prediction of the cosmic core-collapse
rate, based on strong evidence from star formation rate data and direct
confirmation of the minimum stellar mass to lead to a core-collapse
supernova. This is based on observations of living massive stars, and
does not differentiate whether they will die as visible supernovae or
optically-dark collapses.  Since those outcomes have comparable neutrino
emission, this data is enough to precisely predict the DSNB.


\subsection{Measured Supernova Rate}

As noted, the relationship between star formation and supernova rates is
{\it just a simple conversion} with small uncertainties on the result.
Since the redshift evolution of the star formation rate is well
measured, the evolution of the supernova rate is then precisely defined,
and the conversion predicted above could be checked at a single
redshift.

Measuring supernova rate densities is not easy, since supernovae are
infrequent in individual galaxies.  Core-collapse supernovae (Types II,
Ib, Ic) are fainter and less uniform in their optical properties than
the thermonuclear supernovae (Type Ia) used as distance indicators, and
have received much less attention, despite being more common.

At high redshift, repeat imaging of a volume can find supernovae largely
independent of their host galaxies.  This directly measures the quantity
of interest. At low redshift, galaxies from a catalog can be monitored. 
Since this finds the galaxies first, and the supernovae second, it is
subject to incompleteness due to faint galaxies, and one must carefully
define the global volume and time monitored from those for each galaxy. 
While these corrections are made, catalog surveys tend to find lower
supernova rates and less extreme supernovae than volume surveys. 
Lastly, supernova rates estimated using supernovae that were not
collected in a systematic way should be taken only as lower bounds.

\begin{figure}[t]
\centerline{\includegraphics[width=4in]{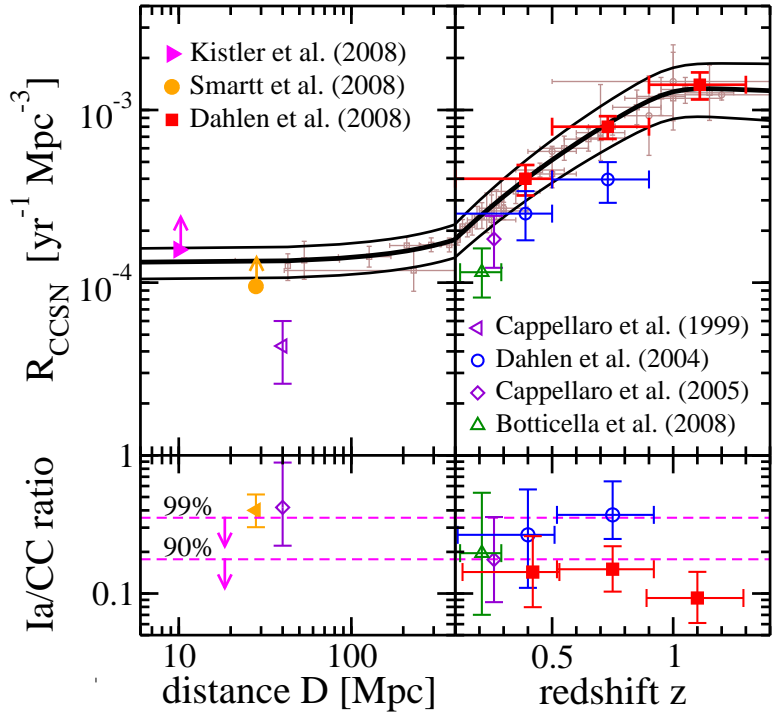}}
\caption{\label{fig:SNrate}
{\bf Predicted and measured cosmic core-collapse supernova rates.}
In the upper panel, the band and its width indicate the absolute
prediction and its uncertainty from the star formation rate data alone
(shown, scaled down, with faint grey points).  The most reliable
supernova rate measurements are shown with solid symbols; those
with open symbols must be less complete.  In the lower panel, rate
ratios are shown, which supports this. Figure as shown in
Ref.~\cite{Horiuchi:2008jz}.
}
\end{figure}

The predicted supernova rate density is shown in {\bf
Figure~\ref{fig:SNrate}}.  There is very good agreement with the most
reliable data, despite the conversion being predicted absolutely, and
not fit. The uncertainty on the star formation rate is taken as $20 \%$
on $R_{SF}(z = 0)$.  The predicted supernova rate and its uncertainty
are $R_{SN}(z = 0) = (1.25 \pm 0.25)
\times 10^{-4}$ Mpc$^{-3}$ year$^{-1}$ (recall Section 3.1). At redshift
zero, the lower limits on the rates are already quite high, matching the
prediction well; the 10 Mpc was defined very conservatively, and this
and the 28 Mpc point are known to be incomplete~\cite{Kistler:2008us,
Smartt:2008zd}. Therefore, supernova rate measurements lower than this
band {\it must be less complete}, since the shape is fixed from the star
formation rate. At high redshift, the updated Dahlen et al.
points are in excellent agreement; these were measured
volumetrically~\cite{Dahlen:2004km, Dahlen08}, and are larger than
points that were not~\cite{1999A&A...351..459C, 2005A&A...430...83C,
2008A&A...479...49B}.
[After Ref.~\cite{Horiuchi:2008jz} was published, the Dahlen et al.
points decreased slightly, but are still consistent with the band shown
in {\bf Figure~\ref{fig:SNrate}} (T. Dahlen, private communication, 2010).]
That faint core-collapse supernovae were
missed is supported by data on their rate relative to the brighter Type
Ia supernovae.  Note that the discovery rates of supernovae have
dramatically increased with time, with advances in technology and
interest.

The agreement of the predicted and measured supernova rates
strongly supports this input for the DSNB calculation.  While a
significant fraction, $\sim 50\%$, of dark collapses is allowed, this is
not required ($\sim 10\%$ is expected). These collapses might be
intrinsically faint, truly dark, or just obscured.  In any case, their
existence can only raise the level of the DSNB.  In the near future,
supernova rate measurements will greatly improve, and their comparison
with star formation rate data will allow more precise tests of the
fraction of dark collapses~\cite{2008ApJ...684.1336K, Horiuchi:2008jz,
2009PhRvL.102w1101L, LFB10}.


\section{Third Ingredient: Neutrino Interactions and Detectors}

The detection probability depends on the product of the DSNB flux and
the effective area of the detector. Since SK (and Earth) is nearly
transparent to neutrinos, the effective area is the neutrino-proton cross
section times the number of target protons, independent of the shape or
orientation of the detector.

While supernovae make all flavors of neutrinos and antineutrinos, the
best prospects are for $\bar{\nu}_e$ in SK, detected by $\bar{\nu}_e + p
\rightarrow e^+ + n$, as it has both large size and low background
rates. The properties of neutrinos are now measured well, and exotic
scenarios that would affect DSNB detection, e.g., large mixing with
sterile species, are no longer favored.


\subsection{Detection of DSNB Electron Antineutrinos}

Electron antineutrinos are detected by the inverse beta decay reaction
on free protons, $\bar{\nu}_e + p \rightarrow e^+ + n$.  Here ``free"
protons means hydrogen nuclei, and not the protons bound in heavier
nuclei, for which nuclear binding effects suppress interactions at low
energies; the atomic ionization state is irrelevant.

The total cross section for inverse beta decay is
\begin{equation}
\sigma(E_\nu) \, = \, 
\left[ 0.0952 \times 10^{-42} {\rm\ cm}^2 \,
(E_\nu - 1.3 {\rm\ MeV})^2 \right] \,
(1 - 7 E_\nu/M_p) \,,
\label{eq:crosssection}
\end{equation}
for neutrino energies above threshold, $E_\nu > 1.8$ MeV, and where
$M_p$ is the proton mass (for $c = 1$). In the high-energy
range, the rising cross section helps offset the falling flux spectrum.
This expression, which is good to the several-percent level in the relevant
energy range, neglects the small positron mass but does include the
few-tens-percent recoil-order corrections, shown outside the square
brackets. At the same orders, the kinematic relationship is
\begin{equation}
E_e \, = \, 
\left[ (E_\nu - 1.3 {\rm\ MeV}) \right]
(1 - E_\nu/M_p) \,,
\label{eq:kinematics}
\end{equation}
where the isotropy of the neutrino angular distribution has been taken
into account. The recoil-order correction to the kinematics can be
neglected relative to the detector energy resolution. Full expressions
for the cross section and kinematics are given in
Ref.~\cite{Vogel:1999zy, Strumia:2003zx}.


\subsection{The Super-Kamiokande Detector}

The fiducial volume of SK contains 22.5 kton of water, and hence $1.5
\times 10^{33}$ free protons~\cite{Fukuda:2002uc}. To reduce
backgrounds, SK is under more than 1 km of rock, is heavily shielded by
separate outer and inner buffer regions, and was constructed to maintain
extreme purity.  SK is large enough to have a few DSNB interactions per
year, and already has low background rates.  These facts are fortuitous,
as the SK design was optimized not for the DSNB, but for measurements of
proton decay, and atmospheric, accelerator, solar, and Milky Way
supernova neutrinos.

In SK, relativistic charged particles are detected by the cones of
optical \v{C}erenkov light produced as they lose energy; the
\v{C}erenkov process is a small component of the energy loss rate, but
is the only one that SK detects directly.  The \v{C}erenkov light
travels through water to the photomultiplier tubes on the walls, which
view a homogeneous volume of transparent water.  From the patterns of
received photons, the position, direction, energy, and identity of
charged particles are determined.  SK has excellent detection efficiency
for the energies considered here.


\section{Framework Redux: DSNB Detection Spectrum at Earth}

Now that we have reviewed the three ingredients -- the supernova
neutrino emission, cosmic supernova rate, and detector capabilities --
we return to the framework for the DSNB detection spectrum at Earth.

The {\it event rate spectrum}, in units s$^{-1}$ MeV$^{-1}$, where the
detection cross section $\sigma(E_\nu)$, the number of proton targets
$N_p$, and the shift between neutrino energy $E_\nu$ and positron
energy $E_e \simeq E_\nu - 1.3$ MeV have been accounted for, is
\begin{equation}
\frac{dN_e}{dE_e}(E_e) \, = \,
N_p \, \sigma(E_\nu) \,
\int_0^\infty
\left[(1+z) \phantom{\frac{a}{b}} \!\!\! \varphi[E_\nu (1 + z)] \right] 
\left[R_{SN}(z) \phantom{\frac{a}{b}} \!\!\! \right]
\left[\left| \frac{c \, dt}{dz} \right| dz\right] \,.
\label{eq:ratespectrum}
\end{equation} 
For relevant energies, those above 10 MeV, the effective upper limit on
the line of sight integral is $z \sim$ 1--2, as beyond there the falling
$dt/dz$ is no longer compensated by a rising $R_{SN}(z)$ and the larger
emission energies required for the same observed energy probe further
into the high-energy tail of the emission spectrum.

\begin{figure}[t]
\centerline{\includegraphics[width=4in]{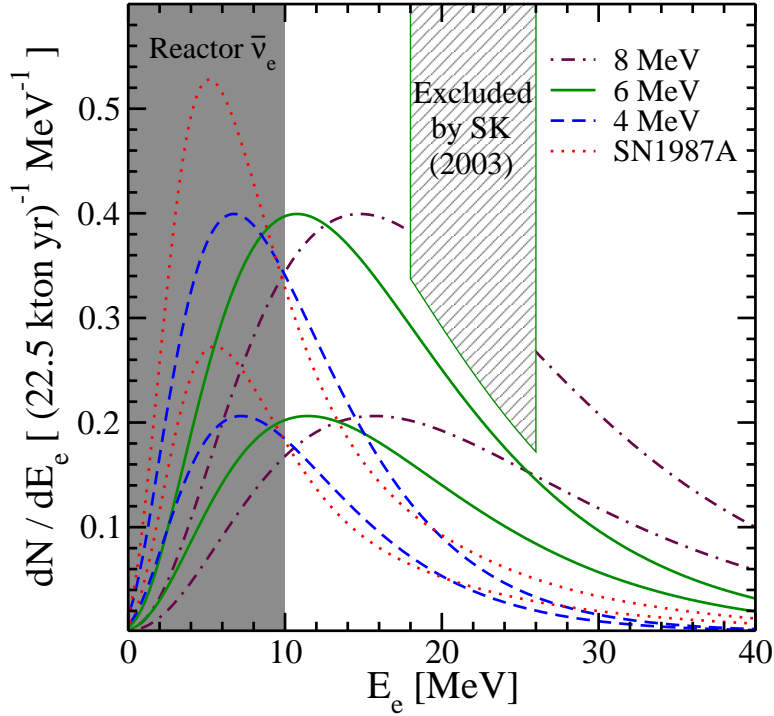}}
\caption{\label{fig:ratespectrum}
{\bf Predicted DSNB $\bar{\nu}_e$ event rate spectrum in positron energy.}
The labeled bands and their widths are as in {\bf
Figure~\ref{fig:fluxspectrum}}. Integrated event rates are tabulated in
Ref.~\cite{Horiuchi:2008jz}. The 2003 exclusion from SK is shown; it is
largely independent of the assumed temperature.  The energy range of
the irreducible reactor background is shaded; backgrounds at higher
energies depend on if gadolinium is added to SK.  Figure as shown in
Ref.~\cite{Horiuchi:2008jz}.
}
\end{figure}

In {\bf Figure~\ref{fig:ratespectrum}}, we show predictions for the DSNB
detection spectrum, taking up-to-date inputs and their uncertainties
into account.  The range due to different emission parameters,
especially at high energies, indicates the uncertainties to be reduced
by measuring the DSNB. The uncertainty on the supernova rate is already
modest and will soon be smaller.  The detection reaction is inverse beta
decay in SK, which is well understood. This predicted event rate
spectrum can be directly compared with measured data, which are a
combination of DSNB signal and detector backgrounds. The DSNB is
especially sensitive to the important high-energy part of the emission
spectrum. The predictions for different temperatures converge near 10
MeV, so a measurement there would probe $E_{\bar{\nu}_e, tot}$ alone;
the falloff of the spectrum would then probe $E_{\bar{\nu}_e, avg}$
alone.

\section{Super-Kamiokande 2003 Upper Limit on the DSNB}

\begin{figure}[t]
\centerline{\includegraphics[width=4in]{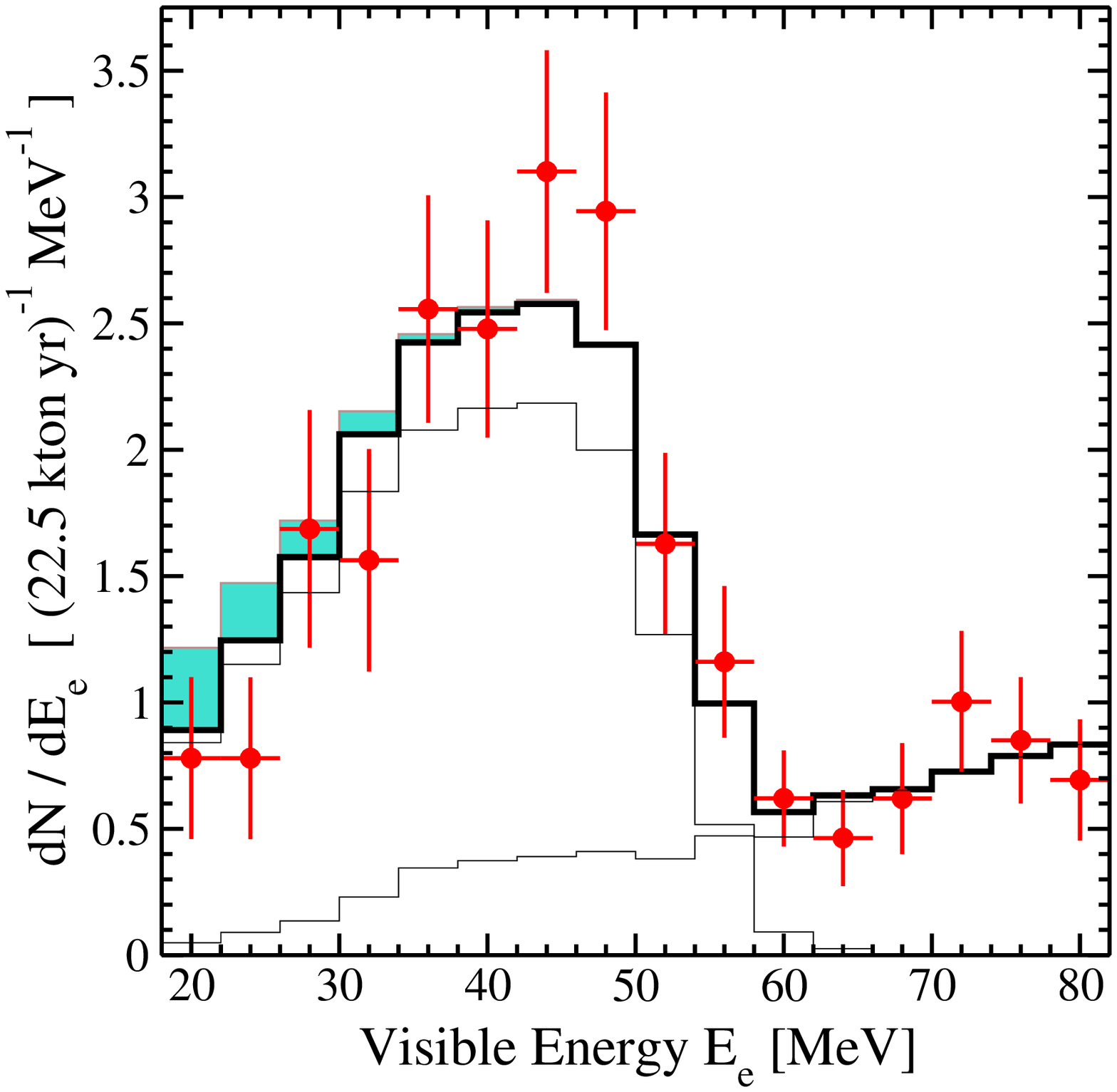}}
\caption{\label{fig:SKrates}
{\bf Results of the SK 2003 DSNB $\bar{\nu}_e$ search.}
The efficiency-corrected measured data are shown with points and error
bars. The expected total atmospheric neutrino background is shown by the
thick solid line, and its components by the thin solid lines. The
largest allowed DSNB signal is shown by the shaded region added to the
atmospheric background. Figure adapted from Ref.~\cite{Malek:2002ns}:
here $dN/dE_e$ is shown, to be directly compared in 1-MeV steps to {\bf
Figure~\ref{fig:ratespectrum}}; to recover the $dN$ results shown in
Ref.~\cite{Malek:2002ns}, multiply the bin values by 4 MeV.
}
\end{figure}

It takes extraordinary efforts to operate a gigantic detector like SK
with low background rates at low energies.  The SK DSNB
limit~\cite{Malek:2002ns} improved upon the Kam-II
limit~\cite{Zhang:1988tv} by a factor $\sim 100$, establishing the only
experimental limit within range of contemporary predictions, an
extremely important achievement.


\subsection{Detector Backgrounds and DSNB Flux Limit}

For the SK 2003 DSNB search, only visible positron or electron (they
cannot be distinguished) energies above 18 MeV were used, where detector
background rates, due to atmospheric neutrinos, are relatively low. 
Nothing beyond expectations was found in this pioneering search,
including any surprise problems that would impede this or improved
future searches.

To understand the atmospheric neutrino backgrounds, it is essential to
use the event rate spectrum in visible energy.  The flux spectrum alone
is misleading, as here the visible energy is quite different than the
neutrino energy, unlike for the DSNB signal.  The backgrounds, shown in
{\bf Figure~\ref{fig:SKrates}} with visible energies of 18--82 MeV,
arise from neutrinos with energies up to $\sim 250$ MeV that invisibly
enter the detector and have charged-current interactions, mostly with
oxygen nuclei, inside the fiducial volume.

Atmospheric $\nu_\mu$ and $\bar{\nu}_\mu$ can produce nonrelativistic
$\mu^-$ and $\mu^+$, which do not produce \v{C}erenkov light as they
lose energy.  When these decay at rest, the electrons and positrons
produced are relativistic, and their spectrum is the bump in {\bf
Figure~\ref{fig:SKrates}}. This is the dominant background, but its
shape is fixed and its normalization can be measured above the energy
range of the DSNB. Atmospheric $\nu_e$ and $\bar{\nu}_e$ can produce
$e^-$ and $e^+$ with energies well below the neutrino energy, due to
nuclear effects. These produce the rise at high energies in {\bf
Figure~\ref{fig:SKrates}}.

A DSNB signal would have appeared as an excess at low energies (compare
to {\bf Figure~\ref{fig:ratespectrum}}); none was seen, and a limit of
$\phi(E_\nu > 19.3 {\rm\ MeV}) \lesssim 1.2$ cm$^{-2}$ s$^{-1}$ at the
$90\%$ confidence level was set~\cite{Malek:2002ns}.


\subsection{Calculated Limits on Supernova Neutrino Emission}

\begin{figure}[t]
\centerline{\includegraphics[width=4in]{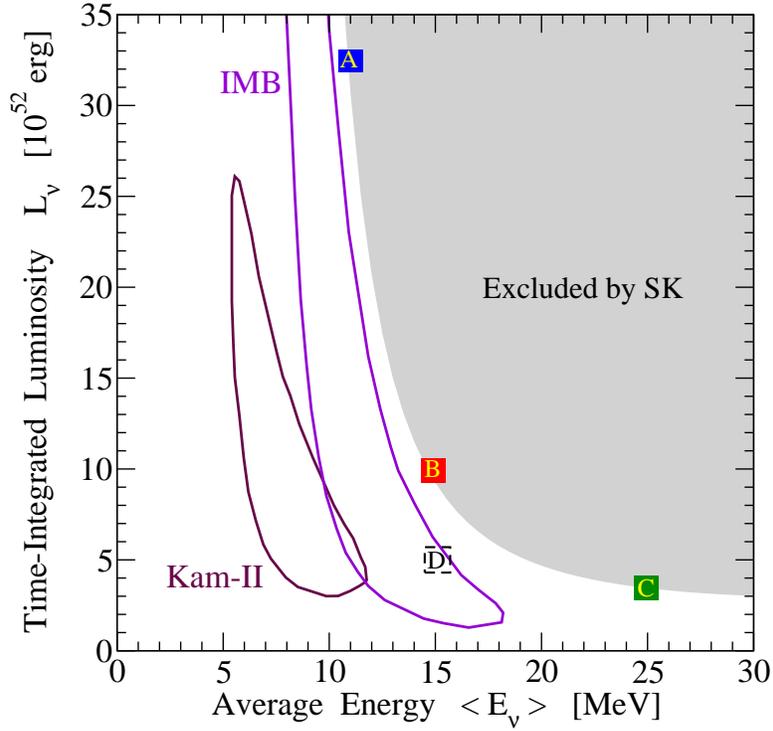}}
\caption{\label{fig:SNparameters}
{\bf Constraints on the effective supernova $\bar{\nu}_e$ emission parameters.}
The grey excluded region follows from the null result of the SK 2003
DSNB search. Approximate regions from fits to the SN 1987A data are
shown, e.g., Ref.~\cite{Jegerlehner:1996kx}. The point D indicates the
nominal parameter values. Figure as shown in Ref.~\cite{Yuksel:2005ae}.
}
\end{figure}

A DSNB limit expressed as a flux integrated above a certain energy,
while a good starting point, is model-dependent and makes comparisons
difficult.  One variable is not enough to characterize DSNB spectra, but
the two natural variables of total energy $E_{\bar{\nu}_e, tot}$ and
average energy $E_{\bar{\nu}_e, avg}$ for the $\bar{\nu}_e$ flavor are
sufficient~\cite{Yuksel:2005ae}.  These directly control our ignorance
of the effective supernova neutrino emission, since the cosmic supernova
rate is now known well enough. This reinterpretation of the SK DSNB
limit, showing how the two emission parameters can compensate each
other, is shown in {\bf Figure~\ref{fig:SNparameters}}.  Any change in
the assumed normalization $R_{SN}(z = 0)$ corresponds to a simple
change in $E_{\bar{\nu}_e, tot}$ alone, and uncertainties on the
evolution $R_{SN}(z) / R_{SN}(z = 0)$ are small.

Excitingly, the sensitivity is near the parameters of supernova
models~\cite{Mezzacappa:2005ju, Ott:2008jb, Janka:2006fh,
Huedepohl:2009wh} and the uncertain fits to the SN 1987A
data~\cite{Jegerlehner:1996kx}. A two-parameter limit on the effective
$\bar{\nu}_e$ emission is simple for experimentalists to implement and
corresponds to the minimal theory.  If it is proven that there is large
variety in supernova neutrino emission, large neutrino mixing effects,
or exotic effects, then appropriate theoretical averages should be
compared to this direct experimental limit.


\section{Pathways to Discovery}

Progress on experimental sensitivity will have a decisive impact, as the
expected signal is close to current limits.  Astronomical observations
and theoretical work will refine and enhance these prospects.

There are two striking features of the SK 2003 results. First, in the
energy range used, the search is background-limited, so the sensitivity
will not improve linearly with exposure time, but naively only with the
square root. Second, going to lower energies would give much more DSNB
signal and much less atmospheric neutrino background; however, there is
a large background rate due to the beta decays of nuclei produced by
spallation interactions of cosmic-ray muons. To detect the DSNB, SK
therefore needs to reduce backgrounds and extend the search to lower
energies.  But how?


\subsection{Proposal for Background Reduction in SK}

Since the DSNB detection reaction is $\bar{\nu}_e + p \rightarrow e^+ +
n$, while the detector background reactions mostly do not produce
neutrons, it is obvious that neutron detection is the key to separating
signal from backgrounds.  The neutron loses energy by elastic
scattering and, at present, captures on a free proton, producing a
2.2-MeV gamma ray that leads to an insufficient detection signal in SK.

The idea of positron-neutron coincidence detection goes back to Reines
and Cowan, and it is well known how to implement this in
oil-and-scintillator detectors, which have much higher light yields than
water-\v{C}erenkov detectors.  For example, with the suspension of a
small concentration of a gadolinium compound in the oil, neutrons will
capture on gadolinium, due to its enormous cross section, producing an
easily detectable 8-MeV gamma-ray signal. The neutron is detected a long
tens of microseconds after the positron, at nearly the same position.

However, until Ref.~\cite{Beacom:2003nk}, it had not been shown that
neutron detection with dissolved gadolinium in a water-\v{C}erenkov
detector could be feasible. The key points are that soluble compounds
exist, e.g., GdCl$_3$ and Gd$_2$(SO$_4$)$_3$, and that the gamma rays
would lead to a detectable signal in SK, due to its high photomultiplier
coverage. That these and many other technical concerns had positive
answers was quite surprising.  The ability to detect neutrons would give
unprecedented capabilities to separate signals and backgrounds in very
large detectors.   As described in Ref.~\cite{Beacom:2003nk}, this
would reduce the atmospheric neutrino backgrounds and, more importantly,
would dramatically reduce the large spallation radioactivity backgrounds,
allowing SK to reduce its DSNB analysis threshold to about 10 MeV, below
which the reactor $\bar{\nu}_e$ background is overwhelming.

\begin{figure}[t]
\centerline{\includegraphics[width=4.0in]{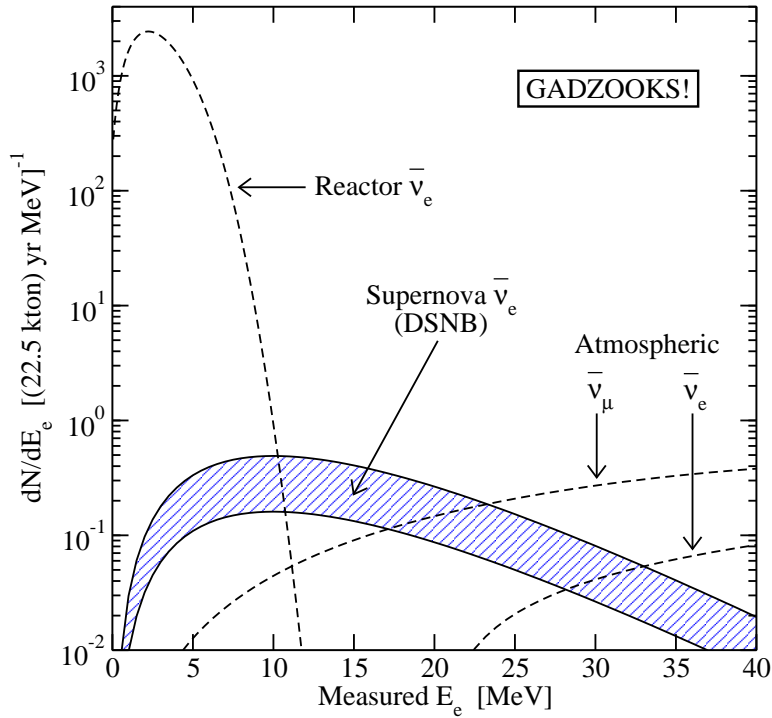}}
\caption{\label{fig:gadzooks}
{\bf Expected detection rates in SK with dissolved gadolinium.}
The DSNB signal could be cleanly detected at a reasonable rate
(see {\bf Figure~\ref{fig:ratespectrum}} for updated DSNB predictions
and uncertainties).  Figure as shown in Ref.~\cite{Beacom:2003nk}.
}
\end{figure}

{\bf Figure~\ref{fig:gadzooks}} shows that there would be excellent
prospects for the DSNB if SK had a $\sim 0.1\%$ concentration of
dissolved gadolinium. There is no known astrophysical source that can
produce a comparable high-energy $\bar{\nu}_e$ flux, leaving a
detection window for the DSNB.  If an exotic process converts a small
fraction of solar $\nu_e$ to $\bar{\nu}_e$, then the discovery of this new
physics could partially get in the way of the DSNB, but we should have
such problems~\cite{Beacom:2003nk, Raffelt:2009mm}.


\subsection{Research and Development for Gadolinium in SK}

From the time of Ref.~\cite{Beacom:2003nk}, Vagins, aided by other SK
experimentalists, has been intensively researching the practicalities
of introducing a dissolved gadolinium compound into
SK, e.g., Refs.~\cite{VaginsGrants1, VaginsGrants2, VaginsGrants3,
Nakahata:2008zz}.  Our initial investigations, and the more extensive
research that followed, have uncovered no insurmountable problems
regarding the infrastructure required to manage the gadolinium, the
effects on the detector and other physics searches, or the
positron-neutron coincidence detection.  However, it has been
determined that it is not advisable to pack a kilogram of white
GdCl$_3$ powder in carry-on luggage for an international flight.

The progress on the comprehensive research and development program, and
the seriousness with which the SK Collaboration is considering adding
gadolinium, can be illustrated with two recent examples.  First, a small
container of gadolinium-loaded water and a neutron source were lowered
into SK, and it was found that the detectable signal of neutron capture
on gadolinium works as expected~\cite{Watanabe:2008ru} (see also
Ref.~\cite{Dazeley:2008xk}). Second, a new 200-ton ``miniature" ($\sim
1\%$ scale) of SK with dissolved gadolinium is being built underground
near SK to test all aspects of a gadolinium-loaded water-\v{C}erenkov
detector~\cite{Kibayashi:2009ih}.


\subsection{Prospects for Other Detectors and Flavors}

Oil-and-scintillator detectors the size of SK would have reduced
atmospheric backgrounds, but the same DSNB signal rate and a
reactor background that is relatively large at any location on
Earth~\cite{Learned:2008zj, Wurm:2007cy}.  Ultimately, a
high-statistics measurement of DSNB $\bar{\nu}_e$ will be needed to
fully develop the science possibilities, probably with a
water-\v{C}erenkov detector at least ten times larger than SK, as is
being considered for other purposes~\cite{Bernstein:2009ms,
Autiero:2007zj}; the prospects for a high-statistics DSNB measurement
have been explored~\cite{Ando:2004sb, Lunardini:2006pd, Volpe:2007qx,
Chakraborty:2008zp, Galais:2009wi}.

Direct DSNB limits on other neutrino flavors are weaker than for
$\bar{\nu}_e$ in SK.  With all their data analyzed, the Sudbury Neutrino
Observatory sensitivity to DSNB $\nu_e$ will only be several times
worse~\cite{Beacom:2005it, Aharmim:2006wq}, comparable to indirect
limits~\cite{Lunardini:2006sn}, but limits for other flavors are worse
by orders of magnitude~\cite{Aglietta:1992yk, Lunardini:2008xd}. Large
liquid-argon detectors are promising for DSNB
$\nu_e$~\cite{Cocco:2004ac, Rubbia:2009md}.


\section{Conclusions}

Neutrino astronomy can revolutionize astrophysics and physics, due to
its unique abilities to see unseen processes and to probe subtle new
particle interactions. Stellar fusion reactions and the core-collapse
mechanism were revealed by the detection of neutrinos from the Sun and
the nearby SN 1987A. Solar neutrinos were crucial to the discovery of
neutrino mixing, and supernova neutrinos probed exotic physics far
beyond the reach of laboratory experiments.

However, the great promise of neutrino astronomy remains largely
unfulfilled, as no other neutrino sources have yet been detected. Recall
the courage of the pioneers of this field, who, more than forty years
ago, had neither adequate detectors nor accurate astrophysics to fortify
their hopes.  We have advantages of many orders of magnitude on each,
and can see the end of the desert.  To make progress on our
understanding of supernovae and their role in astrophysics, as well as
of the neutrinos themselves, we must detect more supernova neutrinos.

\medskip

Fortunately, first detection of the DSNB is within reach. Why are we so
sure?

\smallskip
$\bullet$
{\bf The framework for calculating the DSNB is solid.} \\
It is built on simple principles of cosmology, well tested with photons.

\smallskip
$\bullet$
{\bf Supernova neutrino emission has been measured from SN 1987A.} \\
Theory and indirect evidence suggest it should typically be larger.

\smallskip
$\bullet$
{\bf The uncertainty on the cosmic supernova rate is modest.} \\
Observations of massive stars and supernovae are precise and quickly
improving.

\smallskip
$\bullet$
{\bf All aspects of DSNB signal detection are understood.} \\
Neutrino interactions and detector properties have each been well
measured.

\medskip
Therefore, even including uncertainties, there is a robust prediction
that {\it there must be a few DSNB neutrino events per year in SK}.
These events are hidden by backgrounds -- what about the experimental
prospects for revealing them?

\smallskip
$\bullet$
{\bf SK has already shown sensitivity quite close to these predictions.} \\
Importantly, SK has controlled detector backgrounds nearly to the required level.

\smallskip
$\bullet$
{\bf There are good prospects for improving the SK DSNB sensitivity.} \\ 
With gadolinium, SK would reduce its backgrounds and energy threshold.

\medskip

Therefore, the DSNB could be discovered in SK.  With sufficient detector
background reduction, even a few signal events will be statistically
significant and scientifically important. If the DSNB is not found,
there must be surprising new physics, requiring an upheaval in
astrophysics or exotic new particle physics.

The enduring value of a DSNB measurement will come from its comparison
to other observations and theory, which include the SN 1987A data and
numerical simulations of core collapse. The SK 2003 limit is already an
important new result, as it excludes neutrino emission as large as often
assumed, and is only a factor 2--4 away from more conservative cases.
Significant developments in the theoretical predictions are also eagerly
anticipated, as, after more than forty years of work, recent results
have increasingly realistic physics and accurate calculations, show
explosions in some cases, and some are able to integrate to several seconds.
These models will provide a rich context for interpreting DSNB results and,
eventually, to novel tests of neutrino properties and other new physics.

In the next several years, we will gain a newly deep and comprehensive
understanding of supernovae and their neutrinos, following from the
above improvements converging with unprecedented electromagnetic data,
results from gravitational wave observatories, and, with a little luck,
a Milky Way neutrino burst.


\newpage

{\sc Disclosure Statement} \\
The author is not aware of any affiliations, memberships, funding, or financial
holdings that might be perceived as affecting the objectivity of this review.

\bigskip

{\sc Acknowledgments} \\
The author's research was supported by NSF CAREER Grant PHY-0547102.

I am grateful to my collaborators on the DSNB --
Shin'ichiro Ando, 
Eli Dwek, 
Brian Fields, 
Andrew Hopkins, 
Shunsaku Horiuchi, 
Amy Lien, 
Louie Strigari, 
Mark Vagins, 
Terry Walker, 
Hasan Y\"uksel, 
and 
Pengjie Zhang --
as well as to my additional collaborators on closely-related topics -- 
Dick Boyd, 
Will Farr, 
Joe Formaggio, 
Matt Kistler, 
Chris Kochanek, 
Tony Mezzacappa, 
Jos\'{e} Prieto, 
Matthew Sharp, 
Kris Stanek, 
Yoichiro Suzuki, 
Todd Thompson, 
and 
Petr Vogel -- 
for sharing their knowledge, enthusiasm, and wisdom.

I especially thank Shin'ichiro Ando, Andrew Hopkins, Shunsaku Horiuchi, 
Todd Thompson, Mark Vagins, and Hasan Y\"uksel for helpful comments on
the draft manuscript.



\end{document}